# Structural stability and thermodynamics of CrN magnetic phases from *ab initio* and experiment

Liangcai Zhou[1,*], David Holec[2], Matthias Bartosik[1,4], Fritz Körmann[3], Blazej Grabowski[3], Jörg Neugebauer[3] and Paul H. Mayrhofer[1,4]

[1]*Institute of Materials Science and Technology, Vienna University of Technology, A-1040 Vienna, Austria*

[2]*Department of Physical Metallurgy and Materials Testing, Montanuniversität Leoben, A-8700 Leoben, Austria*

[3]*Max-Planck-Institut für Eisenforschung GmbH, D-40237 Düsseldorf, Germany*

[4]*Christian Doppler Laboratory for Application Oriented Coating Development at the Institute of Materials Science and Technology, Vienna University of Technology, A-1040 Vienna, Austria*

The dynamical and thermodynamic phase stabilities of the stoichiometric compound CrN including different structural and magnetic configurations are comprehensively investigated using a first-principles density-functional-theory (DFT) plus U approach in conjunction with experimental measurements of the thermal expansion. Comparing DFT and DFT+U results with experimental data reveals that the treatment of electron correlations using methods beyond standard DFT is crucial. The non-magnetic face-centered cubic B1-CrN phase is both, elastically and dynamically unstable, even under high pressure, while CrN phases with non-zero local magnetic moments are predicted to be dynamically stable within the framework of the DFT+U scheme. Furthermore, the impact of different treatments for the exchange-correlation (xc)-functional is investigated by carrying out all computations employing the local density approximation and generalized gradient approximation. To address finite-temperature properties, both, magnetic and vibrational contributions to the free energy have been computed employing our recently developed spin-space averaging method. The calculated phase transition temperature between low-temperature antiferromagnetic and high-temperature paramagnetic (PM) CrN variants is in excellent agreement with experimental values and reveals the strong impact of the choice of the xc-functional. The temperature-dependent linear thermal expansion coefficient of CrN is experimentally determined by the wafer curvature method from a reactive magnetron sputter deposited single-phase B1-CrN thin film with dense film morphology. A good agreement is found between experimental and *ab initio* calculated linear thermal expansion coefficients of PM B1-CrN. Other thermodynamic properties, such as the specific heat capacity, have been computed as well and compared to previous experimental data.

1. **INTRODUCTION**

Transition metal nitrides have attracted much interest due to their excellent performance in applications such as hard protective coatings on cutting tools, diffusion barriers, and wear resistant electrical contacts [1-6]. Among the transition metal nitride family, chromium nitride (CrN) is valued especially for its good wear, corrosion and oxidation resistance [3, 4, 6].

There are many experimental reports on CrN due to its wide applicability in the industrial area [3, 7-17]. At room temperature, it adopts a paramagnetic (PM) cubic B1 (NaCl prototype, space group $Fm\bar{3}m$) structure with lattice constant $a$=4.14 Å [12]. Upon cooling below the Néel temperature ($T_N$ = 200-287K) [7, 12, 15, 18], a simultaneous structural and magnetic phase transition to an antiferromagnetic orthorhombic (AFM Ortho, space group $Pnma$) phase induced by magnetic stress takes place, which is accompanied by a discontinuous volume reduction of ~0.59% [7]. The AFM Ortho-CrN phase shows a small structural distortion from the underlying cubic B1 lattice characterized by the angle $\alpha \approx 88.3°$ and the AFM ordering consists of alternating double (110) planes of Cr atoms with spin up and spin down, respectively [12].

Regarding theoretical investigations, CrN has received great attention mainly due to its electronic and magnetic properties [19-26], while only a few papers have been devoted to its phase stability [22, 27]. Most theoretical studies assumed CrN in the non-magnetic rather than in the paramagnetic (PM) state [16, 28-32], while experiments clearly reveal non-vanishing local magnetic moments even above the Néel temperature [33]. The reason for this serious approximation is largely related to the challenges in describing the dynamic magnetic disorder in the paramagnetic state. Recently, Alling *et al*. [22] employed the special quasi-random structures (SQS) approach [34] to simulate the magnetic disorder in PM B1-CrN and concluded that the magnetic disorder together with strong correlation effects in the PM B1-CrN phase largely influence the Gibbs energy. Including the magnetic entropy, a phase transition between the high-temperature PM B1-CrN phase and the low-temperature antiferromagnetic (AFM) ground state with a distorted orthorhombic (Ortho) structure was predicted at 498 K [22]. In their treatment, however, the vibrational contribution to the free energy was not taken into account, despite the known significant influence on thermodynamic properties [35]. Very recently, the same group accomplished to include vibrational contributions via disordered-local-moment molecular-dynamics-simulations in conjunction with the temperature-dependent effective potential (DLM-MD-TDEP) method [36]. The underlying DFT calculations were based on the local density approximation (LDA) with an on-site Coulomb interaction term, U. An improved phase transition temperature of 381 K has been found, which is, however, still ~100 K higher than the experimental observations. Despite the recent progress, many questions are still unanswered. A decisive issue is for instance the impact of the underlying exchange-correlation (xc)-functional on the thermodynamic stability, which is so far not known. Recently, we have developed an alternative method to compute the vibrational Gibbs energy contributions based on a spin-space averaging (SSA) method [37]. Our method does not require computationally demanding molecular dynamics simulations and can be therefore efficiently employed to scan the influence of different xc-functionals as will be performed in the present work.

When discussing phase stability, it is important to distinguish energetic (meta)stability, mechanical and dynamical stability. The first one is related to the Gibbs free energy, *G*, which predicts the phase with the lowest *G* to be the most stable one, all other phases being metastable or even instable. Regarding mechanical stability the Born-Huang criteria [38] for elastic constants have to be fulfilled. The dynamical stability considers the complete vibrational spectrum of a material; a material is (dynamically) stable when no imaginary phonon frequencies exist. In other words, the Born-Huang mechanical stability criteria *provide* a *necessary* condition for the dynamical stability, but not a *sufficient* one. For instance, by explicitly computing phonon dispersions, imaginary frequencies can occur at the Brillouin zone boundary (indicating dynamical instability), even if the Born-Huang criteria are fulfilled [39]. It is therefore necessary to perform explicit phonon calculations to evaluate the dynamical stability. Such an evaluation of the dynamical stability is computationally much more expensive as calculating the full tensor of elastic constants (in particular for systems with large numbers of atoms in their unit cells). Consequently, explicit phonon computations have been so far mostly neglected and we therefore present here a complete study of CrN phase stabilities.

The nonmagnetic (NM) B1-CrN phase has been reported to be energetically unfavorable with respect to other spin arrangements with non-zero local magnetic moments [22]. It is however not clear whether NM B1-CrN is dynamically unstable with respect to mechanical distortions, i.e. whether the elastic constants fulfill criteria for mechanical stability [38] and/or if there are any imaginary frequencies in its phonon dispersion. Another question of interest is whether the NM B1-CrN can be stabilized, e.g. by high pressure, and thus represents a potential metastable state. The literature data based on density functional theory (DFT) suggest its mechanical stability [28, 29, 31, 32], but an explicit evaluation via the full phonon spectrum analysis will be presented for the first time in the present paper.

As mentioned above, another important and so far unresolved issue is the impact of strong electron correlation effects on the stability of CrN phases. For example, it has been argued that LDA+U is more suitable than the generalized gradient approximation (GGA) [22], but a comprehensive study employing GGA+U, which might turn out to be even more appropriate, is so far missing. Additionally, the impact of the previously proposed value for the on-site Coulomb interaction parameter U on the vibrational free energy has not been discussed so far as well [36].

The aims of the present paper can be hence summarized as follows: (1) To comprehensively investigate the energetic, mechanical and dynamical stability of CrN including non-magnetic, anti-ferromagnetic, ferromagnetic and paramagnetic configurations, (2) to elucidate the impact of the chosen xc-functional approximation and *U*-parameter, and (3) to compute the thermodynamic stability of the different variants. Firstly, elastic constants are calculated and used to evaluate the mechanical stability based on the Born-Huang mechanical stability criteria. As discussed above, these stability criteria might be not sufficient. Therefore, in a second step, phonon calculations are performed to prove the dynamical stability. Eventually, Gibbs free energies based on the SSA formalism [37] including vibrational and magnetic contributions are computed to predict the phase transition in CrN employing GGA, GGA+U and LDA+U approaches, and the thus obtained transition temperatures are compared with available experimental results. Furthermore, thermodynamic properties such as lattice thermal expansion coefficient and specific heat capacity of the high-temperature PM B1-CrN are presented and, where applicable, compared to experimental data. In particular, we compare our DFT computed linear thermal expansion coefficient with a set of newly determined experimental data obtained by the wafer curvature method and from a single-phase face-centered cubic B1-CrN thin film.

II. **Methods**

**A. Static DFT calculations**

As discussed above, unlike other early transition metal nitrides, CrN requires extra efforts due to the non-zero local magnetic moments of the Cr atoms. In this work the SQS approach [22, 34, 40] is used to model the PM state by employing 2x2x2 cubic supercells with 32 Cr and 32 N atoms and randomly distributing the spin-up and spin-down moments on the Cr-atoms. In order to provide a complete picture of phase stabilities, we consider in the present work also B1-CrN with ferromagnetic (FM) and AFM ordering consisting of alternating single (001) planes of Cr atoms with spin up and spin down in addition to the NM B1, PM B1 and AFM Ortho structures of CrN (cf. the crystal structures in Fig. 1).

All first-principles calculations are based on DFT as implemented in the Vienna *Ab initio* Simulation Package (VASP) [41, 42]. The ion-electron interactions are described by the projector augmented wave method (PAW) [43] with a plane wave energy cutoff of 500 eV. The semi-core *p* states are treated as valence for Cr ($3p^6 3d^5 4s^1$) while there are 5 valence electrons for N ($2s^2 2p^3$). In order to take into account the strong on-site Coulomb interaction (U) caused by the localized 3*d* electrons of Cr, the local density approximation (LDA) and the generalized gradient approximation (GGA) plus a Hubbard U-term method is adopted within the framework of the Dudarev formulation [44, 45]. Here, only the difference U-J, with J being the screened exchange energy, determines the material properties. Alling *et al.* [22] tested U-J in the range from 0 to 6 eV in their previous LDA+U study, and found U-J = 3 eV to yield an optimal description of the structural and electronic properties of AFM and PM-CrN. Apart from structural and electronic properties, the inclusion of the U-parameter is also decisive for the correct description of the magnetism in CrN. This is exemplified in Fig. 1, where we show that a too small value of the U-J parameter might even yield a wrong magnetic ground state for CrN. The total energy was evaluated for a number of structures as function of U-J ranging from 0 to 6 eV. For the sake of clarity, only a few structures are listed. It becomes obvious that both, standard LDA and GGA-PBE [46] predict the NM Bh-CrN phase (TaN prototype, space group *C2mm*) to be the ground state of CrN, which disagrees with experimental observations of AFM Ortho-CrN [7]. The energy difference between AFM Ortho-CrN and AFM B1-CrN becomes smaller as the U-J value increases. Our results suggest that the value of U-J should be larger than 1.5 eV (LDA) or 0.25 eV (GGA) in order to obtain AFM Ortho-CrN as the ground state. Following the results of Alling *et al.* [22], we adopted the value of U-J=3 eV also for our DFT+U calculations. In addition, the conventional GGA-PBE (without the +U term) is used in the present work to demonstrate the impact of the on-site Coulomb interaction on the vibrational properties.

The energy convergence criterion for electronic self-consistency was set to 0.1 meV/atom. The Monkhorst-Pack scheme [47] was used to construct *k*-meshes of 3x3x3 (128-atom supercells), 3x4x5 (96-atom supercells), 5x5x5 (64-atom supercells), 12x12x12 (8-atom cells), and 21x21x21 (2-atom cells). The elastic constants were calculated using the stress-strain approach discussed in detail in our previous work [40].

### B. Phonon calculations

The phonon calculations for NM, FM and AFM B1-CrN were performed employing 2x2x2 supercells consisting of 64 atoms constructed from a conventional face-centered cubic cell with 8 atoms. For AFM Ortho-CrN, 2x3x2 supercells with 96 atoms were created from its unit cell with 8 atoms. Super cell size convergence tests were performed for NM and FM B1-CrN using up to 4x4x4 supercells (based on primitive cells with 2 atoms) containing 128 atoms. The 64-atoms SQS adopted to simulate the paramagnetic state in B1-CrN was also used for the corresponding phonon calculations. Note that the computation of vibrational properties for paramagnetic materials is a tremendous task due to the delicate coupling of magnetic and atomic degrees of freedom [48]. We use here our recently developed statistical spin space averaging (SSA) procedure [37], which allows the computation of forces in paramagnetic materials. Within the SSA approach, the SSA force on an atom *j* is given as the gradient on the SSA free energy surface $F^{SSA}$ as [37]:

$$F_j^{SSA} = -\frac{\partial F^{SSA}}{\partial R_j} = \sum_m p_m F_j^{HF}\{(R_j),\sigma_m\}, \qquad (1)$$

where $F_j^{HF}\{(R_j),\sigma_m\}$ are the Hellmann-Feynman forces for an individual magnetic configuration $\sigma_m$ and $p_m = \exp[-\frac{E^{B0}\{(R_j),\sigma_m\}}{k_B T}]$ denotes the Boltzmann weights. As discussed in Ref. [37], the summation over different magnetic configurations, $\sigma_m$, in the above equation can be expressed by a sum over lattice symmetry operations. In the present work we employ such a summation over lattice-symmetry equivalent forces. After applying the symmetry operations, these forces correspond to locally inequivalent magnetic configurations and allow one to perform the SSA procedure based on a single magnetic SQS structure. In the PM regime it is typically sufficient to restrict the sum in Eq. (1) to completely disordered configurations, since they dominate the partition sum (i.e. the weights $p_m$). Usually about 50-100 magnetic configurations yield converged forces [37]. Note that the treatment of magnetic partially disordered configurations requires advanced sampling techniques, which are beyond the scope of the present manuscript and will be discussed elsewhere [49]. In the present work the magnetic disordered configurations $m$ are constructed by all 3N Cartesian (positive and negative) displacements for the given SQS super-cell, resulting in 384 magnetic configurations.

For the purpose of comparison, we additionally employed the *conventional* method for computing PM phonons [50], i.e. we allow atomic relaxations in the PM state which results in virtual displacements at T=0 K. The consequences of this approach and the necessity to employ the SSA averaging technique will be discussed later. Also the impact of choosing a different SQS will be discussed. The finite-displacement method implemented in the phonopy [51] combined with VASP was used to calculate the real-space force constants and corresponding phonon properties. The residual forces (background forces) in the unperturbed super cell were subtracted from the force sets of the displaced structures. The summation over lattice symmetry equivalent forces according to the SSA scheme [37] has been carried out by phonopy software [52]. In order to benchmark the direct-force constant method, the NM B1-CrN, FM B1-CrN, AFM B1-CrN and AFM Ortho-CrN real-space force constants were additionally calculated within the density functional perturbation theory framework [53] as implemented in the VASP code. The phonon frequencies were subsequently calculated using the phonopy code. The phonon densities of states (DOS) from both methods (direct force constant and perturbation theory) coincide without any noticeable differences (0.5 meV/atom and 2 meV/atom difference in the free energy evaluated using both methods at 0K and 1000K, respectively). Anharmonic phonon-phonon interactions which can become important close to the melting point [54] are not relevant for the present study. We will show this explicitly in Sec. C by comparing our calculations with the molecular dynamics simulations performed in Ref. [36], which implicitly include anharmonicity.

When LDA+U or GGA+U is used to treat the strong correlation effects, a small band gap opens for AFM Ortho-CrN and PM B1-CrN. Consequently, the dipole–dipole interactions including properties of the Born effective charge tensor and dielectric tensor, which result in a longitudinal-transversal optical phonon branches (LO-TO) splitting [53], should be taken into account during the phonon calculations. For that purpose the dipole-dipole interactions were calculated from the linear response method within the density functional perturbation theory framework as implemented in VASP at the $\Gamma$-point of reciprocal space. The contribution of non-analytical term corrections to the dynamical matrix developed by Wang *et al*. was considered by the following formula [55]:

$$\Phi_{\alpha\beta}^{jk}(M,P) = \phi_{\alpha\beta}^{jk}(M,P) + \frac{1}{N}\frac{4\pi e^2}{V}\frac{[qZ^*(j)]_\alpha [qZ^*(k)]_\beta}{q\varepsilon_\infty q}, \qquad (2)$$

where $\phi_{\alpha\beta}^{jk}$ is the contribution from short-range interactions based on super-cell, $N$ is the number of primitive unit cells in the super cell, $V$ is the volume of the primitive unit cell, $q$ is the wave vector, $\alpha$ and $\beta$ are the Cartesian axes, $Z^*(j)$ is the Born effective charge tensor of the $j^{th}$ atom in the primitive unit cell, and $\varepsilon_\infty$ is the high frequency static dielectric tensor, i.e., the contribution to the dielectric permittivity tensor from the electronic polarization.

### C. Thermodynamic properties

Once the phonon DOS is obtained, the vibrational energy and its effect on thermal properties can be directly evaluated. In combination with first-principles calculations, the Helmholtz free energy, $F(V,T)$, is the most convenient choice for a thermodynamic potential, since it is a natural function of *V* and *T*:

$$F(V,T) = E_0(V) + F^{el}(V,T) + F^{vib}(V,T) + F^{mag}(V,T), \qquad (3)$$

where $E_0(V)$ is the internal energy at 0 K obtained from the equation of state [56], $F^{el}(V,T)$ and $F^{vib}(V,T)$ are the thermal electronic and lattice contributions to the free energy, respectively. Further details on the *ab initio* calculations of $F^{el}(V,T)$ and $F^{vib}(V,T)$ are discussed in Ref. [57] and references therein. $F^{mag}(V,T)$ is the magnetic free energy which is in the present work considered within the mean-field approximation as [58]:

$$F^{mag} \approx -k_B T \ln(M(T,V)+1),  \qquad (4)$$

where $M(T,V)$ is the magnitude of the local magnetic moment (in units of $\mu_B$) and $k_B$ is the Boltzmann constant. In this work we used the averaged magnitude of the local magnetic moment, $M(T=0\ K, V=V_0)$, at 0 K at the ground state volume, $V_0$, to predict the magnetic entropy. Our test calculations revealed that the inclusion of the volume dependence of $M(T,V)$ has no significant impact on our main results and conclusions, and will be therefore not considered in the following. The average magnetic moments based on GGA, GGA+U, and LDA+U schemes, which are in the following used in Eq. (4), are 2.48 $\mu_B$, 2.94 $\mu_B$, and 2.83 $\mu_B$, respectively, which agrees well with previous theoretical values [22, 27].

### D. Experimental details

Since no experimental study has been devoted so far to the lattice thermal expansion coefficient, $\alpha$, this property has been measured within the present work. A single-phase face-centered cubic CrN thin film was deposited on Si (100) substrates (7 x 21 mm$^2$) using reactive magnetron sputtering. The deposition was carried out at a deposition temperature of 743 K in an Ar and N$_2$ gas atmosphere of a total pressure of 0.4 Pa and a constant Ar/N$_2$ flow ratio of 2/3. A target power of 250 W and a 3" Cr target (purity 99.9 %) were used. During the deposition, a bias potential of −70 V was applied to the substrates to ensure dense film morphology. The film thickness was measured using cross-sectional scanning electron microscopy. The biaxial stress in CrN was recorded as a function of temperature using the wafer-curvature method [59]. The wafer-curvature system was operating with an array of parallel laser beams and a position sensitive charge-coupled device detector. The sample was heated by a ceramic heating plate with a constant heating rate of 5 K/min from room temperature to 518 K under vacuum conditions of ≤ 10$^{-4}$ mbar. The maximum temperature was chosen to be clearly below the deposition temperature to avoid thermally activated processes in the film (e.g. recovery of deposition-induced defects) and the substrate and thus to guarantee pure thermoelastic behavior. The film stress was deduced from the sample curvature 1/$R$ according to:

$$\sigma = \frac{Mh^2}{6Rt_f} , \qquad (5)$$

with $M \sim 180$ GPa being the biaxial modulus of the substrate, $h = 380$ μm the substrate and $t_f = 1.50$ μm the film thicknesses. Due to the mismatch in the coefficients of thermal expansion of CrN and Si, the temperature change results in the formation of thermal stresses in the film. When the elastic moduli are taken as constants in the given temperature range, the linear thermal expansion coefficient, $\alpha$, of CrN thin film is related to that of substrate (Si) as:

$$\alpha(T) = \alpha_{si(001)}(T) + \frac{1-\nu_{CrN}}{E_{CrN}} \cdot \frac{d\sigma}{dT} , \qquad (6)$$

with $E_{CrN}$=330 GPa, $\nu_{CrN}$=0.22 [60, 61] and the thermal expansion coefficient of substrate (Si) is expressed as [62]:

$$\alpha_{si(001)}(T) = (3.725[1-\exp(-5.88\times10^{-3}[T-124]+5.548\times10^{-4}T)\,])\times10^{-6}\,(T\text{ in K}). \qquad (7)$$

## III Results and Discussion

### A. Elastic constants and mechanical stability

The elastic constants of the different studied structures as calculated by the GGA, GGA+U, and LDA+U schemes are listed in Table I. It can be seen that most of the elastic constants derived from GGA and GGA+U are smaller than those obtained from the LDA+U scheme. This is consistent with the fact that GGA frequently overestimates lattice parameters leading to an underestimation of binding energies, elastic properties, and phonon frequencies for most materials [63]. In contrast, LDA (or LDA+U) tends to overestimate binding as expressed by smaller lattice parameters, and higher elastic constants and phonon frequencies when compared with experimental or GGA and GGA+U values. A comparison with the limited available experimental data reveals good agreement for the $C_{11}$ component (in particular for GGA+U), while $C_{12}$ and $C_{44}$ seem to be overestimated by all the LDA+U, GGA and GGA+U theoretical predictions. However, it might be that the substantially too low experimental $C_{ij}$ constants are a consequence of the fact that the measurements have been performed on a polycrystalline sample at room temperature, i.e., the measured elastic constants may be influenced by soft grain boundaries [61].

The mechanical stability of any crystal requires the strain energy to be positive, which implies that the whole set of elastic constants, $C_{ij}$, must satisfy the Born-Huang stability criterion

[38]. Using this requirement Table I shows that CrN is mechanically stable within the GGA scheme for all magnetic and structural configurations studied here in consistency with previous results [28, 29, 32]. In contrast, we observe that $C_{44}$ derived from the LDA+U and GGA+U schemes suggest NM B1-CrN to be mechanically unstable ($C_{44} < 0$). This result underlines the importance of strong correlation effects, which were not considered in previous elastic constant calculations using standard LDA or GGA approaches. Finally, we observe AFM B1-CrN to show a small tetragonal distortion, which is reflected in the number of nonequivalent elastic constants: $C_{11}$, $C_{12}$, $C_{13}$, $C_{33}$, $C_{44}$ and $C_{66}$ in agreement with previous results [28].

### B. Dynamical stability

As discussed above, a necessary condition for a structure to be dynamically stable is that it is stable against all possible small perturbations of its atomic structure, i.e., that all phonon frequencies must be real. The GGA, GGA+U, and LDA+U phonon spectra of NM B1-CrN are presented in Fig. 2 (a). Imaginary acoustic branches at the *X* and *W*-points of the Brillouin zone suggest an internal instability of NM B1-CrN. These phonon anomalies are due to the high electron DOS at the Fermi level of NM B1-CrN [19], which causes the existence of soft phonon modes at these points. We evaluated the phonon density of states (DOS) of NM B1-CrN for several pressures up to 900 GPa to check whether applying external pressure can eliminate these imaginary phonons. For the sake of clarity, only the representative LDA+U results are presented in Fig. 2 (b), yielding the following tendency: NM B1-CrN is dynamically unstable even under high pressures, indicating the NM B1-CrN cannot be stabilized by high pressure.

In order to check the phase stability of CrN for various magnetic states, the phonon spectra of CrN from GGA, GGA+U, and LDA+U with ordered magnetic states are presented in Fig. 3. The comparison of the phonon spectra of FM B1-CrN, AFM B1-CrN and AFM Ortho-CrN from GGA, and GGA+U or LDA+U reveals that the phonon spectra are significantly shifted to higher frequencies for the latter approximations. The GGA and GGA+U based phonon frequencies are always lower than those derived from the LDA+U scheme, a trend already reflected by the softer elastic constants when GGA or GGA+U is used. For FM and AFM B1-CrN, when no U-correction is used, a softening tendency for the phonon branch around the *X*-point is observed, and for AFM B1-CrN phonon frequencies even decrease to negative values. The dynamical instability, not observed via the elastic constants calculations, indicates once more the necessity of explicit phonon calculations for the stability analysis. This phonon anomaly observed for the GGA calculations originates from the high electron DOS at the Fermi level. When the on-site Coulomb repulsion U in the DFT+U scheme is switched on, the electron density of states at Fermi level significantly decreases for the FM B1-CrN although no gap opens [21]. Here, one should note that despite the fact that the DFT+U approximation does not result in a gap opening at the Fermi level for NM- and FM B1-CrN, as shown in Refs. [19, 21], it significantly increases the stability of FM and AFM B1-CrN as measured by phonon frequencies around *X*-point. No imaginary frequencies in the phonon dispersion curves are observed independent of whether the GGA+U or LDA+U scheme is used, implying that FM B1-CrN, AFM B1-CrN and AFM Ortho-CrN are all dynamically stable, and represent hence potential (meta)stable phases.

In contrast to the ordered magnetic states of FM B1-CrN, AFM B1-CrN and AFM Ortho-CrN, the magnetic state in PM B1-CrN is disordered. As a first step we allow for atomic relaxations at 0 K. Since at finite temperatures the magnetic fluctuations are usually faster compared to the atomic motion, these displacements can be considered as artificial. In order to test the effect of such artificial static displacements, the atomic positions are fixed to the ideal B1 sites and cell shape remains cubic (we refer to this structure as "unrelaxed" hereafter). The phonon DOS corresponding to unrelaxed and fully relaxed SQS are presented in Fig. 4 (a). Clearly, the phonon DOS of the unrelaxed SQS exhibits some different features than that of the fully relaxed one, which means that the artificial static displacements have some impact on the phonon calculations. A closer inspection of the phonon dispersion of the unrelaxed structure reveals some imaginary phonon modes at the $\Gamma$ point. In order to elucidate if these imaginary frequencies are physical, we employ in the following the recently developed SSA method, Eq. (1), to compute the phonon frequencies in the paramagnetic regime. In fact, no imaginary phonon frequencies are obtained (see the red dashed line in Fig. 5) revealing the applicability of the method and the fact that the imaginary phonon modes along the acoustic dispersions are indeed caused by the artificial $T$=0 K relaxations. In order to further analyze the impact of artificial $T$=0 K relaxations on the specific SQS structure, we have performed similar calculations for a second SQS with a different spin arrangement. Figure 4 (b) shows that the phonon DOS's for the two SQS almost fully coincide, suggesting that a particular spin arrangement and consequently a specific spin-induced relaxation of atoms at $T$=0 K in the SQS of PM B1-CrN does not significantly affect the phonon DOS and hence thermodynamic properties. Since the SSA computed phonon DOS does not show any imaginary frequencies and does not require any (unphysical) $T=0$ K relaxations, we will proceed with the SSA obtained phonon DOS and thermodynamic properties for the following thermodynamic analysis. As discussed above, super cell convergence tests for NM and AFM variants have revealed that a 2x2x2 super-cell is sufficient for an accurate description within the considered temperature range. This is in agreement with the findings of Shulumba *et al.* [36], who reported convergence with respect to the super cell size for PM B1-CrN for a super cell with 64 atoms (2x2x2), which we will employ in the following.

Figure 5 presents the phonon spectra of PM B1-CrN derived from GGA, GGA+U and LDA+U schemes employing the SSA method, and in comparison with available experimental data (open circles [23] ) at the $\Gamma$ point. It demonstrates that the transverse branches derived from DFT+U are significantly shifted to higher values and the LO-TO splitting happens at the $\Gamma$ point, when comparing to the result from GGA scheme. The obtained dispersion from DFT+U is in excellent agreement with experimental data (blue circles), especially the results derived from the GGA+U scheme are very close to experimental data. The accurately predicted transverse and longitudinal optical phonon frequencies at the $\Gamma$ point further validate the reliability of the SSA scheme. The phonon spectra derived from GGA, GGA+U, and LDA+U in combination with the SSA scheme confirm that PM B1-CrN is dynamically stable.

C. **Thermodynamic stability and thermodynamic properties of CrN**

The Gibbs free energy has been evaluated according to the treatment described in Sec. C employing in particular the SSA method for the PM phases. Due to the opening of a small gap in AFM Ortho-CrN and PM B1-CrN as obtained from GGA+U and LDA+U [24], the thermal electronic contribution $F^{\text{el}}(V,T)$ to the Helmholtz free energy can be ignored for AFM Ortho-CrN and PM B1-CrN.

Firstly, we compare the SSA results for the Gibbs energies to previous results from the DLM-MD-TDEP method [36] to verify the reliability of our method and also to check the influence of anharmonic effects on the vibrational free energy. The comparison between the present work and the results from DLM-MD-TDEP method [36] is presented in Fig. 6. As the latter method is based on MD simulations, anharmonic contributions are implicitly included. This allows us to evaluate the importance of anharmonic effects for the considered temperature regime. Figure 6 demonstrates that the vibrational free energies of PM B1-CrN from both methods (SSA and DLM-MD-TDEP) fully coincide with each other and that there is no noticeable difference even at high temperatures. The largest deviation of about 5 meV/atom is observed for AFM Ortho-CrN at 1000 K. Keeping in mind that the phase transition under consideration occurs at room temperature, we obtain an excellent agreement with the previous work in the relevant temperature regime revealing that anharmonic contributions are not decisive for the considered transition.

Next we concentrate on the impact of the xc-functional. The differences of harmonic vibrational free energies as a function of temperature between AFM Ortho-CrN and PM B1-CrN in the GGA, GGA+U and LDA+U schemes are presented in Fig. 7 (a). For all considered scenarios, the vibrational contribution to the free energy favors the PM B1-CrN phase. This contribution becomes smaller when the strong correlation effects are taken into account by using Hubbard Coulomb term U.

We now include the magnetic entropy, Eq. (4), into our free energy computations. The results including both, vibrational as well as the magnetic entropy terms evaluated using GGA, GGA+U and LDA+U schemes, are presented in Fig. 7 (b) as a function of temperature for AFM Ortho-CrN and PM B1-CrN phases. The critical temperature for the structural transition between AFM Ortho and PM B1 CrN is $T$s=428 K for GGA, $T$s=300 K for GGA+U, and $T$s=370 K for the LDA+U scheme. The excellent agreement of the latter with the previous theoretical results [36] reveals once more the reliability of our SSA approach. A particularly important conclusion is that the GGA+U value turns out to be significantly closer to the experimental value compared to the previous LDA+U estimations. This improvement of GGA+U compared to the LDA+U data is remarkable and of similar magnitude and importance as the inclusion of vibrational contributions [22,36].

The inclusion of vibrational free energy contributions is considerable and shifts the transition temperature, e.g. for GGA, by almost 500 K [22]. The inclusion of magnetic entropy turns out to be similarly important. To elucidate the impact of magnetic contributions, we included in Fig. 7 (b) also the Gibbs free energy curves *without* the magnetic entropy term $S^{mag}$ in Eq. (4) (dotted lines). Consequently, AFM Ortho-CrN becomes energetically preferred over the PM B1-CrN phase in the whole investigated temperature range. In agreement with previous works [22, 36] we can therefore conclude that both, magnetic and vibrational contributions are equally important to accurately predict the transition temperature. The remaining difference between predicted and experimental values for the phase transition temperature may be related to the approximate treatment of magnetism (neglected non-collinear magnetic configurations, magnetic entropy extracted from mean-field approximation) or to the inherent DFT approximation, i.e. the treatment of the xc-functional.

The phase transition from AFM Ortho-CrN to PM B1-CrN induced by temperature can be alternatively triggered by pressure. In order to investigate the pressure effect on the phase transition, the Gibbs energy is computed via the Legendre transformation as

$$G(p,T) = F(V(p),T) + pV(p,T) \qquad (8)$$

where $F$ is the Helmholtz free energy, Eq. (3), evaluated within the framework of the quasi-harmonic approach at several volumes including the magnetic entropy contributions from Eq. (4) for PM B1-CrN. $p$ is the pressure and $V(p,T)$ is the volume as a function of pressure and temperature. Using the Gibbs energies of AFM Ortho-CrN and PM B1-CrN based on the GGA+U and LDA+U schemes, the $p$-$T$ phase diagram was derived (Fig. 8). The transition temperature increases with pressure in agreement with experimental and theoretical results reported in literature. The transition temperature at $p$=0 GPa obtained from experiment [7, 12, 15] varies from 200 K to 287 K, showing an uncertainty of up to 87 K. Within this error-bar, our theoretical predictions based on the GGA+U method are in excellent agreement with the experimental work and yielding a significant improvement as compared with the previously reported theoretical results. It should be noted that the transition temperatures at $p$=0 GPa are approximately 10 K lower than the results in Fig. 7. This is due to the fact that the free energy in Eq. (3) is evaluated within the quasi-harmonic approximation, while the results presented in Fig. 7 are obtained from the harmonic approach. This implies that the quasi-harmonic contribution (i.e., caused by thermal expansion) is not as important as for example the strong correlation effect or the magnetic entropy contribution, within the considered temperature-pressure regime.

On the other hand, thermal expansion is a key thermodynamic property of PM B1-CrN when it is applied as a high temperature coating material. If compared to experiment, the derivative of the linear thermal expansion coefficient, $\alpha(T)$, provides an even more sensitive measurement of the accuracy achievable by the theoretical predictions. The calculated $\alpha(T)$ of PM B1-CrN is compared to our experimental data in Fig. 9. The results are in good agreement around room temperature, and the discrepancy increases with raising temperature. It was recently demonstrated that $\alpha(T)$ of CrN strongly depends on the grains size [64]. Depending on the thin film microstructure, $\alpha(T)$ varies in the range of $6.7\times10^{-6}$/K and $9.8\times10^{-6}$/K, which for this material somewhat complicates a distinct comparison with our predictions. Considering these complications, the overall agreement with experiment is very reasonable.

Finally, in Fig. 10 we plot the specific heat capacity, $C_P(T)$, at constant (ambient) pressure as a function of temperature predicted from our GGA+U and LDA+U calculations, together with available experimental data from literature [65]. The GGA+U values are slightly closer to the experimental data. This is consistent with the findings for the phase transition temperature discussed above; however, the differences between LDA+U and GGA+U are only minor. The literature data exhibit a peak near the phase transition temperature (~280 K), due to the Néel transition, which has not been accounted for in the present study. This peak originates from the magnetic phase transition and can be resolved only when the magnetic contribution is exactly evaluated by, for example the Heisenberg model for spin interactions [66]. Nevertheless, our predictions are reasonably accurate from 0K to about 200K and also above 600K.

## IV. Conclusions

The phase stability of different structural and magnetic configurations of stoichiometric CrN is studied systematically by first-principles calculations based on the GGA, GGA+U, and LDA+U schemes. In combination with our recently developed SSA procedure, the phonon contributions in paramagnetic materials are computed. A comparison of the three exchange-correlation approximations demonstrates, that strong correlation effects have a significant impact on the mechanical and phase stability of CrN. The elastic constants and phonon spectra show that the non-magnetic B1-CrN phase is dynamically unstable even under high pressures, due to the high electron density of states at Fermi level. The (meta)stability of the ferromagnetic and antiferromagnetic B1-CrN phase is significantly improved when strong correlation effects are considered using the DFT+U approach.

Including the vibrational, electronic and magnetic free energy contributions, the results of our LDA+U-SSA based approach agree well with previous LDA+U-MD simulations. By performing for the first time finite-temperature GGA+U simulations for CrN we show that the treatment of the xc-functional, i.e. GGA+U vs. LDA+U, is decisive to predict accurate phase transition temperatures in CrN. In particular we find that GGA+U significantly improves previous LDA+U predictions. The phase transition between AFM Ortho-CrN and the paramagnetic B1-CrN phase is predicted to be 293 K at ambient pressure being in excellent agreement with the experimental value of 200-287 K. The impact of the xc-functional is similar in magnitude to the impact of vibrational contributions to the phase stability. The linear thermal expansion coefficient, $\alpha(T)$, and the heat capacity, $C_p$, of PM B1-CrN as a function of temperature is obtained from experimental measurements and *ab initio* calculations. The comparison between the experimental results and predictions for these thermodynamic properties reveals good agreement and further confirms the reliability of our theoretical method.

## ACKNOWLEDGEMENTS

The financial support by the START Program (Y371) of the Austrian Science Fund (FWF) and the Austrian Federal Ministry of Economy, Family and Youth and the National Foundation for Research, Technology and Development is gratefully acknowledged. Funding by the European Research Council under the EU`s 7th Framework Programme (FP7/2007-2013)/ERC Grant agreement 290998 and by the collaborative research center SFB 761 "Stahl - ab initio" of the Deutsche Forschungsgemeinschaft is also gratefully acknowledged. First-principles calculations were carried out partially on the cluster supported by the Computational Materials Design department at the Max-Planck-Institut für Eisenforschung GmbH in Düsseldorf and the Vienna Scientific Cluster (VSC). The authors are thankful to Dr. Yuji Ikeda from Kyoto University for valuable help with the phonopy software.

Table I. $T$=0 K elastic constants in GPa of CrN within the framework of GGA, GGA+U, and LDA+U methods including a comparison with experimental values.

| Structure | | C11 | C22 | C33 | C12 | C13 | C23 | C44 | C55 | C66 |
|---|---|---|---|---|---|---|---|---|---|---|
| NM B1 | GGA | 580 | | | 210 | | | 8 | | |
| | GGA+U | 477 | | | 266 | | | -120 | | |
| | LDA+U | 641 | | | 260 | | | -59 | | |
| FM B1 | GGA | 348 | | | 117 | | | 74 | | |
| | GGA+U | 508 | | | 108 | | | 156 | | |
| | LDA+U | 589 | | | 128 | | | 162 | | |
| AFM B1 | GGA | 535 | 535 | 567 | 126 | 86 | 86 | 150 | 94 | 94 |
| | GGA+U | 555 | 555 | 389 | 98 | 66 | 66 | 166 | 129 | 129 |
| | LDA+U | 696 | 696 | 722 | 113 | 76 | 76 | 174 | 124 | 124 |
| PM B1 | GGA | 516 | | | 115 | | | 116 | | |
| | GGA+U | 538 | | | 88 | | | 143 | | |
| | LDA+U | 649 | | | 99 | | | 145 | | |
| | Exp. [60, 61] | 540 | | | 27 | | | 88 | | |
| AFM Ortho | GGA | 439 | 529 | 495 | 195 | 110 | 114 | 221 | 125 | 103 |
| | GGA+U | 444 | 524 | 497 | 169 | 84 | 92 | 223 | 151 | 137 |
| | LDA+U | 503 | 580 | 626 | 228 | 87 | 102 | 275 | 155 | 137 |

Figure captions

FIG. 1. (Color online) $T$=0 K total energy of the stoichiometric compound CrN in different structural and magnetic configurations as a function of the U-J term in the (a) LDA+U and (b) GGA+U schemes. The energy of the AFM Ortho-CrN phase is used as reference. The vertical dash-dotted line indicates the U-J used in the present work. The crystal structures to the right show the various investigated atomic and magnetic arrangements with the blue (white) balls indicating Cr (N) atoms.

FIG. 2. (Color online) (a) Phonon spectra of NM B1-CrN calculated using GGA, GGA+U, and LDA+U, and (b) phonon DOS as a function of pressure for NM B1-CrN from the LDA+U scheme. The gray shaded region highlights imaginary frequencies.

FIG. 3. (Color online) Phonon spectra of (a) FM B1-CrN, (b) AFM B1-CrN and (c) AFM Ortho-CrN calculated using the GGA, GGA+U and LDA+U schemes.

FIG. 4. (Color online) Phonon DOS's from the LDA+U scheme (a) for fully relaxed and unrelaxed SQS, (b) for two SQSs with different SRO parameters.

FIG. 5. (Color online) Phonon spectra of PM B1-CrN simulated with a 64-atom SQS from the GGA, GGA+U and LDA+U schemes in combination with SSA approach, together with experiments at the $\Gamma$ point from Raman and infrared measurements (open circles [23] ).

FIG. 6. (Color online) Comparison of the vibrational free energies for AFM Ortho-CrN (blue lines) and PM B1-CrN (black lines) under the framework of LDA+U. The solid lines denote the results from SSA approach in the present work and dotted lines denote DLM-MD-TDEP method [36].

FIG. 7. (Color online) (a) The differences of vibrational free energy between AFM Ortho-CrN and PM B1-CrN in the GGA, GGA+U and LDA+U schemes, and (b) calculated phase transition temperature from Helmholtz free energies of AFM Ortho-CrN and PM B1-CrN by taking lattice vibrational and magnetic contributions into account under the framework of GGA and LDA+U scheme. The dotted lines denote the Helmholtz free energies of PM B1-CrN without magnetic contributions.

FIG. 8. (Color online) Calculated pressure-temperature phase diagram of CrN based on the GGA+U (black line) and LDA+U (blue lines) schemes compared to experiment (open circles [12] and squares [15]). The solid lines represent the phase diagram based on the relevant spectrum of excitations: total energy, vibrational and magnetic entropy contributions . For LDA+U, the dashed line shows the phase diagram if only the total energy and magnetic excitations are included to emphasize the impact of vibrational entropy.

FIG. 9. (Color online) Linear thermal expansion coefficient, $\alpha(T)$, of PM B1-CrN computed within the GGA+U (black solid line) and LDA+U (blue dashed line) schemes compared to experimental data (open squares) as obtained in the present work.

FIG. 10. (Color online) Heat capacities of PM B1-CrN in the GGA+U (black solid line) and LDA+U (blue dashed line) schemes compared to experimental values (open triangles [65]).

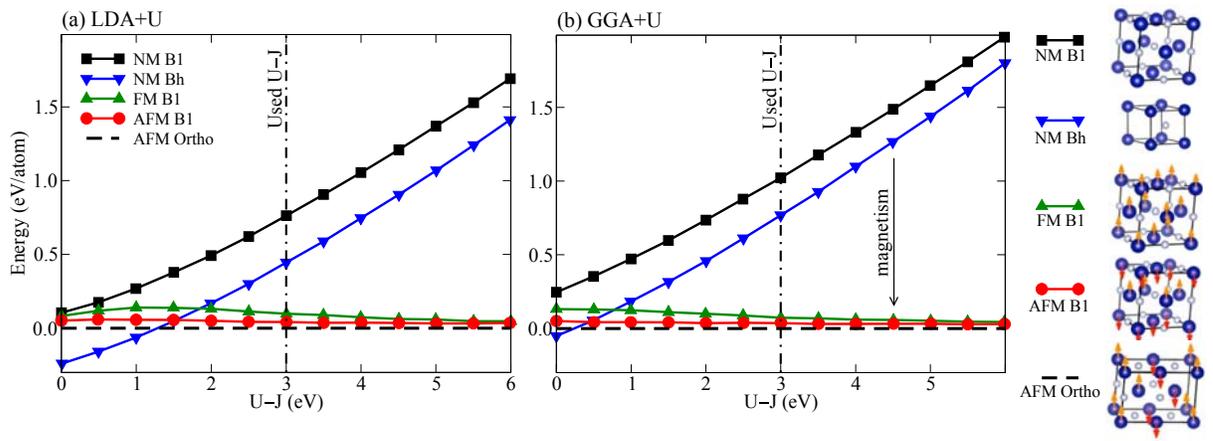

FIG. 1

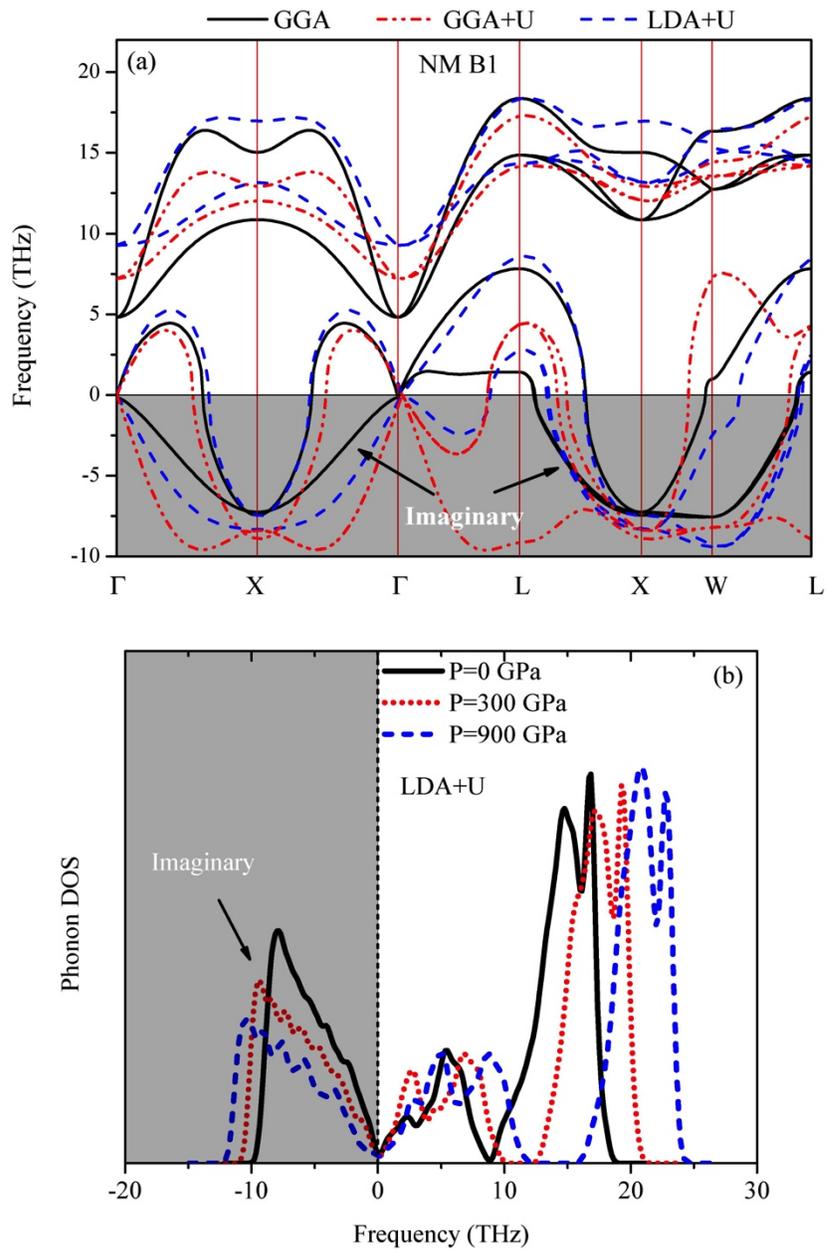

FIG. 2

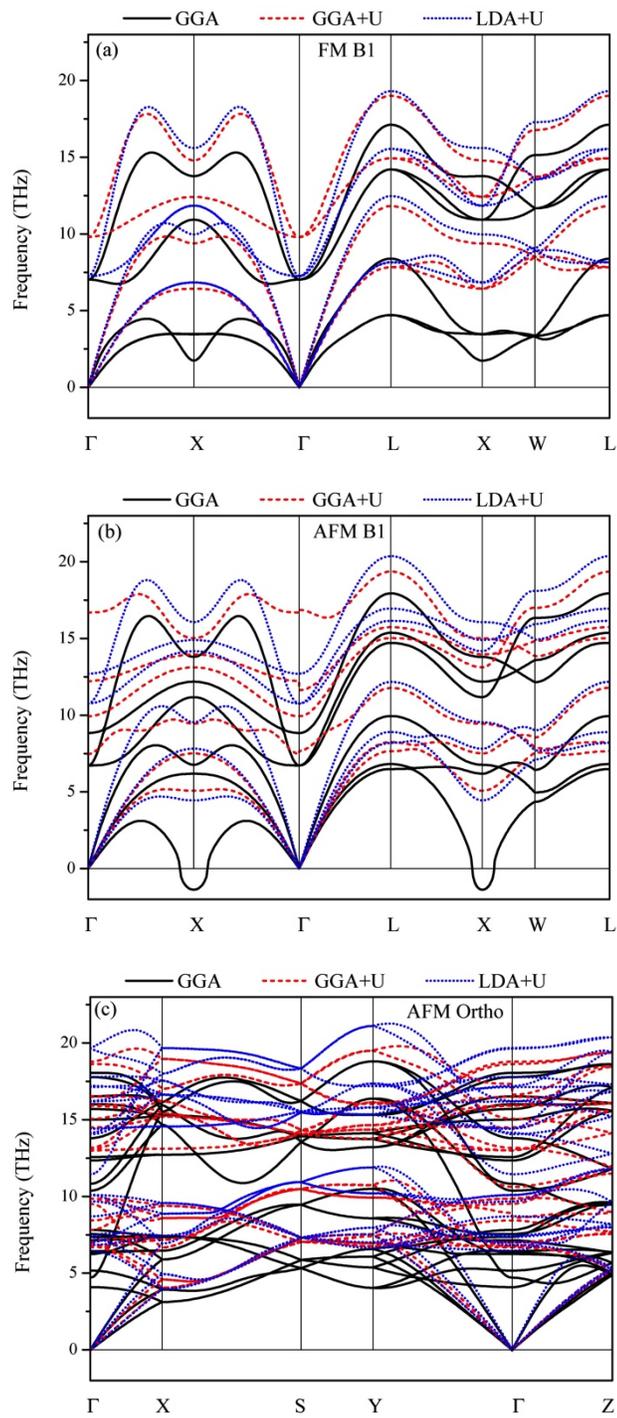

FIG. 3

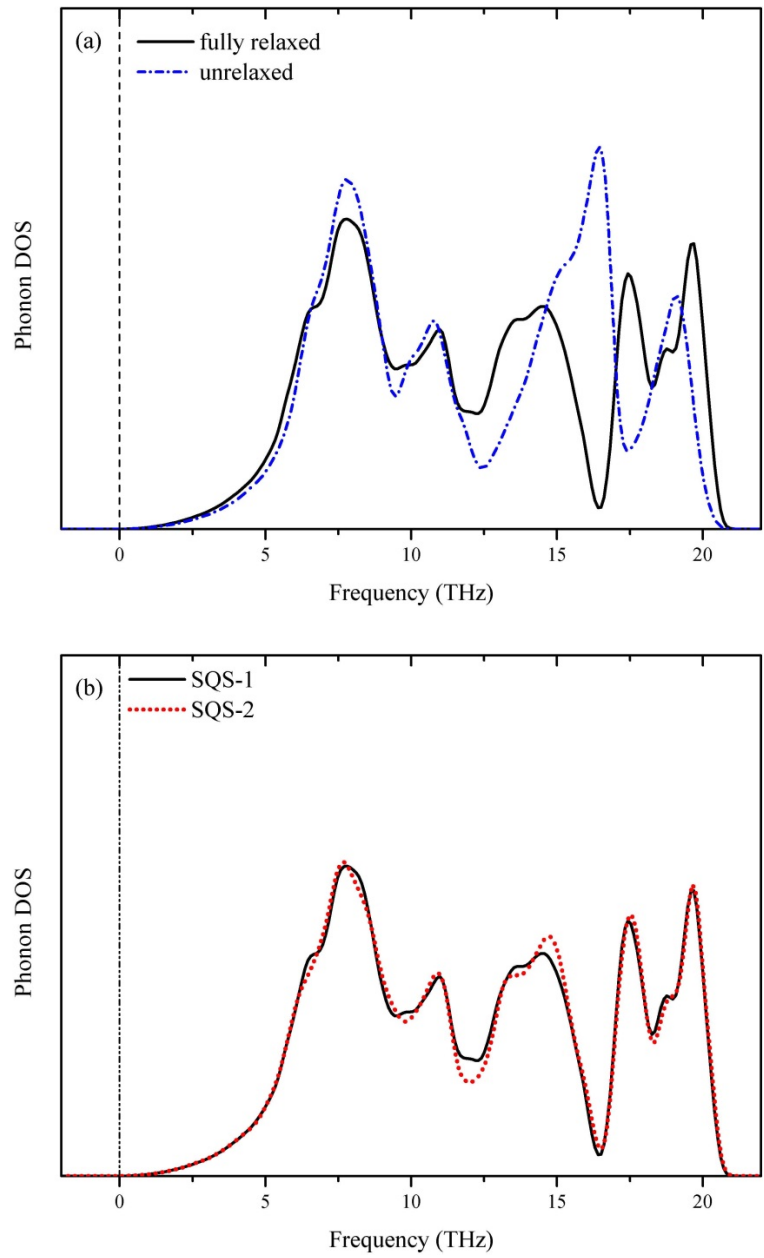

FIG. 4

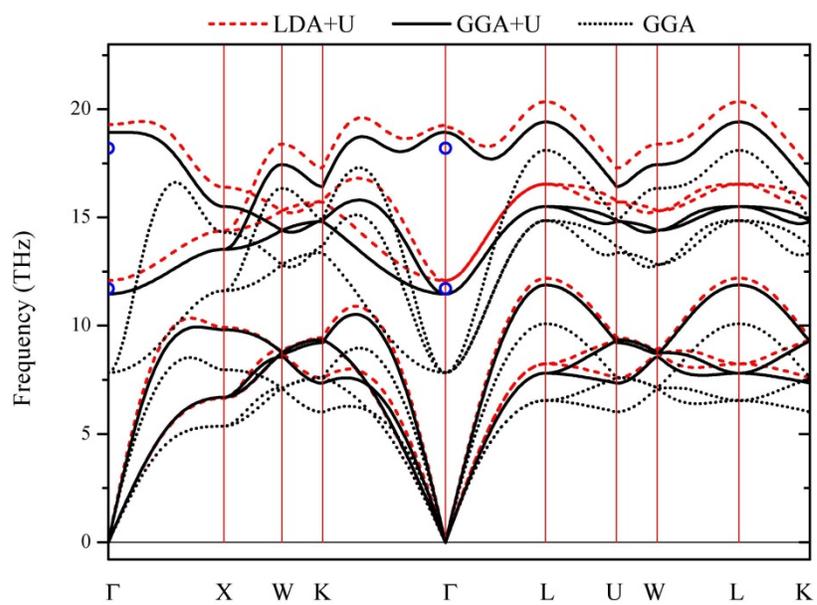

FIG. 5

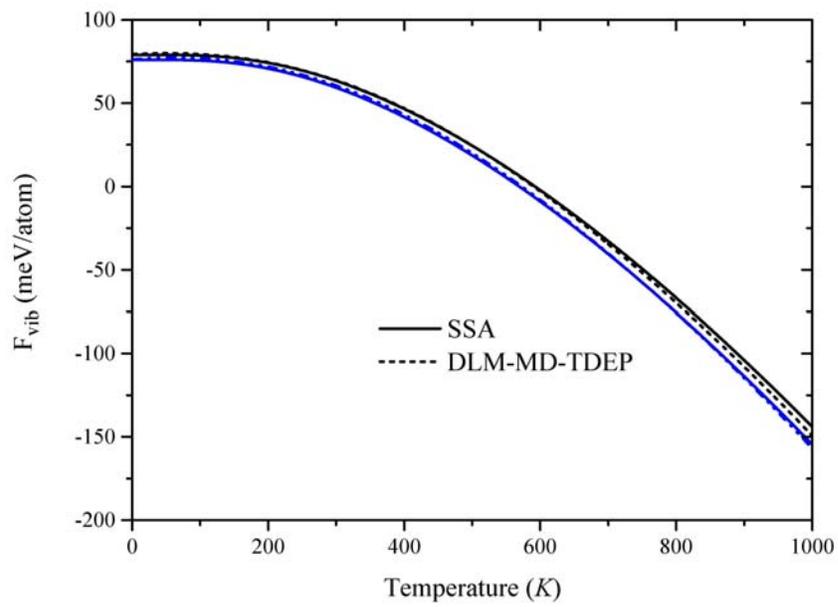

FIG. 6

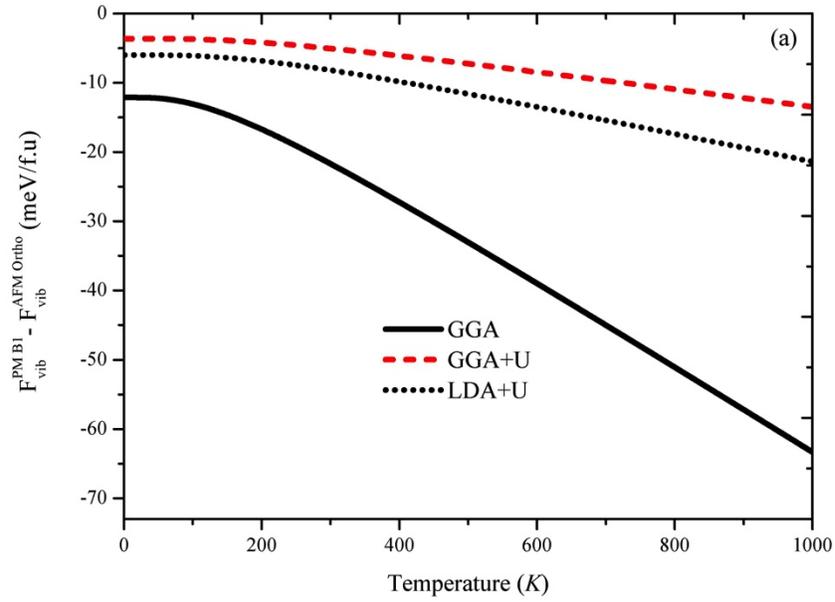

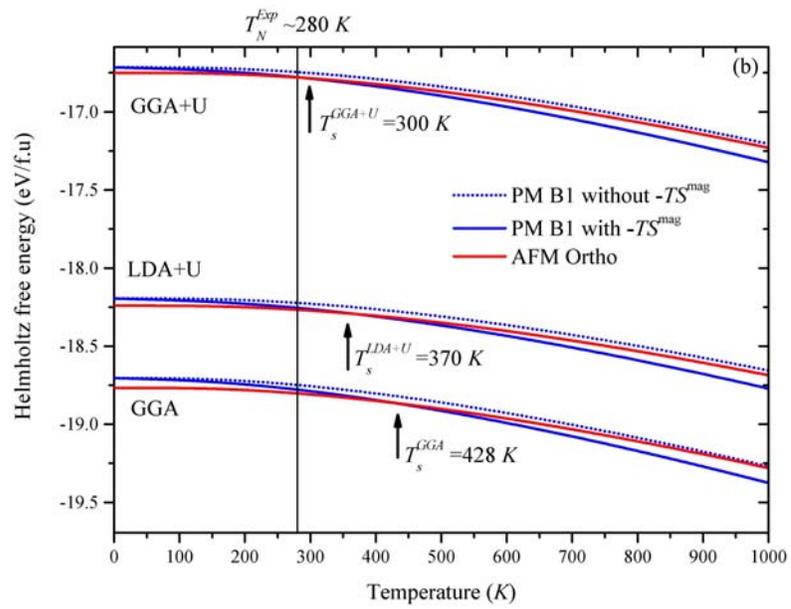

FIG. 7

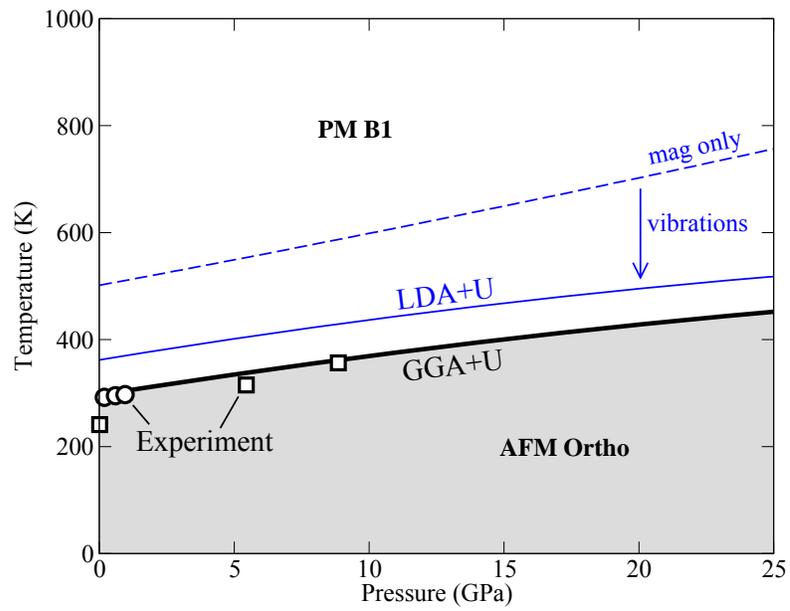

FIG. 8

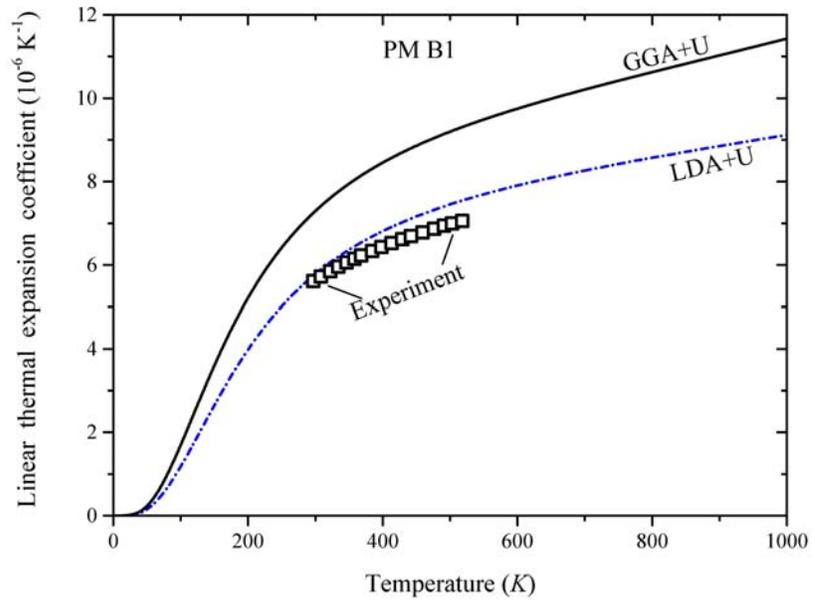

FIG. 9

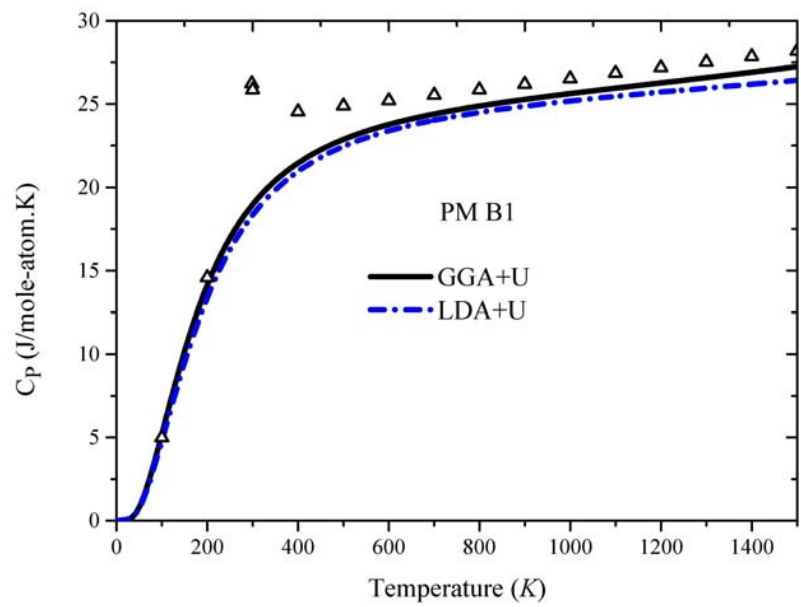

FIG. 10